\documentclass[doublecol]{epl2} 
\usepackage{hyperref}
\usepackage{float}
\usepackage{bbm}
\usepackage[dvipsnames]{xcolor}
\usepackage{bbold}
\def\I{\openone}
\def\openone{\mathbb I}
\usepackage{amsmath}
\usepackage{multirow}
\usepackage{amsfonts}
\usepackage{amssymb}
\usepackage{placeins}
\usepackage{cuted}
\usepackage{flushend}

\usepackage{multicol}

\title{Spin-half fermions endowed with bosonic traces --- Towards phases and classes}
\shorttitle{Spin-half fermions endowed with bosonic traces} %Insert here a short version of the title if it exceeds 70 characters

\author{Rodolfo Jos\'e Bueno Rogerio\inst{1}}
\shortauthor{Rodolfo Jos\'e Bueno Rogerio}

\institute{                    
  \inst{1} Instituto de F\'isica e Qu\'imica, Universidade Federal de Itajub\'a - IFQ/UNIFEI, \\
Av. BPS 1303, CEP 37500-903, Itajub\'a - MG, Brazil.
}

\abstract{
This article focuses on investigating all the possible choices of the phase factor present in the quite recent (unconventional) spin-half particles \cite{dharamboson}, looking towards fully defining the necessary criteria to classify them within an appropriated classification \cite{rodolfospinhalf}. Consequently, we provide the first details on the singular spin-half fermions with bosonic traces}

\begin{document}

\maketitle

\section{Introduction}\label{intro}
One of the most important building blocks in modern physics stands for the spinor fields. We highlight that historically the pioneer case was the Dirac spinor field. And later it has been recognized that Dirac spinors are only a small part of a varied population of spinorial fields \cite{lounestolivro}. The Dirac spinor, belonging to the class of regular spinors in the lights of Lounesto classification \cite{lounestolivro}, can in fact be defined as the spinor whose scalar and pseudo-scalar bilinear quantities are not both identically equal to zero. However, this leaves the door open for an altogether different type of spinors having both scalar and pseudo-scalar bilinear quantities vanishing identically, called singular.

The class of regular spinors is composed of 3 classes of spinors, which in turn, are (naively) called by Dirac spinors. Of these three classes, only one of them, in fact, is genuinely composed of Dirac spinors, with a specific dynamics that ensures a local theory. This class stands for the Lounesto class 2 \cite{rjfermionicfield}. These observations were first noticed by D. V. Ahluwalia, thus, some of the results highlighted here, come from a private communication between the author and him, more details are given in Refs \cite{aaca, mdobook}. Singular spinor fields, may be unusual but they still have a quite interesting physical content to be yet explored \cite{dharamnewfermions,mdobook}. In fact, singular spinors can be further split into three sub-classes: the flag-dipole, flag-pole and dipole spinors. In which the well-known examples stands for the flag-pole (Majorana) and dipole (Weyl) spinors. 

Despite the success of the Lounesto classification, physics is in constant progress, and therefore, some new entities that have been theoretically proposed \cite{dharamboson,dharamnewfermions,mdobook}, may not be accommodated in any of the six classes of Lounesto. Thus, making explicit the need to revisit it in order to fully accommodate these new spinors \cite{beyondlounesto,rodolfospinhalf}. The purpose of this construction is to show the correct form of the bilinear structures, taking into account the bosonic representation of Clifford's algebra, in order to provide correct physical information, present in each of the classes of these new spinors.
As can be seen in \cite{dharamboson}, quite recently it has been proposed a new perspective pointing towards the unification of fermions and bosons. A hint of the physical significance of the new spinors lays in the cosmic structure formation and also in star and galaxy formation, moreover, they may contribute zero to the cosmological constant \cite{dharamboson}. 

This paper deal with a mathematical device that turns possible to right determine the classes
of the new spin-half particles belonging to the classification presented in Ref \cite{rodolfospinhalf}.  The present mathematical apparatus scrutinize all the possibilities of the arbitrary phase factor carried by such new spin-half particles, belonging to class-2 within the quite recent classification shown in \cite{rodolfospinhalf}, also providing a local theory. We may display separately the results for regular and singular spin-half fermions with bosonic traces. We expect that theories defined upon such spinors carry bosonic traces like those presented in \cite{dharamboson}.

The paper is organized as follows: In the next section we delve into a brief overview about spinorial structures and we took the opportunity to revisit the new proposed classification and the bosonic representation of the Clifford algebra. Going further, we deep analyse the mathematical conditions to ascertain the three classes of regular and singular new spinors and we also investigate some of its consequences. Finally, we conclude.

\section{Elementary review}\label{prelude}
This section is reserved for a brief review on some introductory elements, necessary for the study carried along the present paper.

\subsection{Setting the notation}\label{conceitos}
To obtain an explicit form of a given $\psi(\textbf{p})$ spinor we first call out attention for the rest spinors, $\psi(\textbf{0})$. For an arbitrary momentum we have the following condition
\begin{equation}\label{1}
\psi(\textbf{p}) = e^{i\kappa.\varphi}\psi(\textbf{0}),
\end{equation}
where the $\psi(\textbf{0})$ rest frame spinor is a direct sum of the $(1/2,0)$ and $(0,1/2)$ Weyl spinors, which usually is defined as 
\begin{equation}
\psi(\textbf{0}) = \left(\begin{array}{c}
\phi_R(\textbf{0}) \\ 
\phi_L(\textbf{0})
\end{array} \right),
\end{equation}
and the $(1/2, 0)\oplus(0, 1/2)$ boost operator reads
\begin{eqnarray}\label{boostoperator}
\kappa = \sqrt{\frac{E+m}{2m}}\left(\begin{array}{cc}
\I + \frac{\vec{\sigma}\cdot\vec{\textbf{p}}}{E+m} & 0 \\ 
0 & \I - \frac{\vec{\sigma}\cdot\vec{\textbf{p}}}{E+m}
\end{array} \right).
\end{eqnarray} 

Thus, in this section we call attention for some details concerning a quite new particles introduced in \cite{dharamboson}, standing for fermions with bosonic features which reside in the $(1/2, 0) \oplus (0, 1/2)$ representation space, up to constant multiplicative factors of the type $\alpha\times e^{i\beta}$ where $\alpha\in\mathbb{C}$ and $\beta$  $\in \mathbb{R}$ --- the demand of locality forces one to choose $\beta=0$ \cite{dharamboson} --- therefore, such (regular) structures may be written in the following fashion\footnote{We will use the notation $\alpha_{R} $ and $\alpha_{S}$ to distinguish the phase factors from the regular and the singular spinors.} 
{
\footnotesize
\begin{eqnarray}
\lambda_1(\textbf{0})=\sqrt{\frac{m}{2}}\left(\!\!\!\begin{array}{c}
\alpha_{R} \left(\!\!\!\begin{array}{c}
-1 \\ 
i
\end{array}\!\!\!\right) \\\\ 
\quad\left(\!\!\!\begin{array}{c}
i \\ 
1
\end{array}\!\!\!\right)
\end{array}\!\!\!\right),\;\lambda_2(\textbf{0})=\sqrt{\frac{m}{2}}\left(\!\!\!\begin{array}{c}
\alpha_{R}\left(\!\!\!\begin{array}{c}
1 \\ 
i
\end{array}\!\!\!\right) \\ \\
\quad\left(\!\!\!\begin{array}{c}
-i \\ 
1
\end{array}\!\!\!\right)
\end{array}\!\!\!\right),\label{set-1}\nonumber\\
\end{eqnarray}
}
and
{\footnotesize
\begin{eqnarray}
\lambda_3(\textbf{0})=\sqrt{\frac{m}{2}}\left(\!\!\!\begin{array}{c}
\alpha_{R}\left(\!\!\!\begin{array}{c}
1 \\ 
-i
\end{array}\!\!\!\right) \\\\ 
\quad\left(\!\!\!\begin{array}{c}
i \\ 
1
\end{array}\!\!\!\right)
\end{array}\!\!\!\right),\;\lambda_4(\textbf{0})=\sqrt{\frac{m}{2}}\left(\!\!\!\begin{array}{c}
\alpha_{R}\left(\!\!\!\begin{array}{c}
-1 \\ 
-i
\end{array}\!\!\!\right)\\\\ 
\quad\left(\!\!\!\begin{array}{c}
-i \\ 
1
\end{array}\!\!\!\right)
\end{array}\!\!\!\right).\label{set-2}\nonumber\\
\end{eqnarray}
}
Defined for an arbitrary momentum $\lambda_{j}(\textbf{p})=\kappa\lambda_{j}(\textbf{0})$. Therefore, the introduced spinors stand for spinors with locality phases previously defined in \cite{dharamboson}, and the focus of the present work point towards unveiling relevant information regarding regular and also singular classes within a new spinor classification, which may be defined upon the right observance of the parameter $\alpha_{R}$.

\subsection{The new spinor classification}\label{subLounesto}
Let $\lambda$ be an arbitrary spinor field, belonging to a section of the vector bundle $\mathbf{P}_{Spin^{e}_{1,3}}(\mathcal{M})\times\, _{\rho}\mathbb{C}^4$, where $\rho$ stands for the entire representation space $D^{(1/2,0)}\oplus D^{(0,1/2)}$. Thus, the bilinear covariants associated to $\lambda$ yield \cite{rodolfospinhalf} 
\begin{eqnarray}
&&\sigma_{B} = \lambda^{\dag}a_0\lambda,  \nonumber\\
 &&\omega_{B} = i\lambda^{\dag}a_0 a_5\lambda,\nonumber\\
\label{bosonbilinear} &&\textbf{J}_{B}= \lambda^{\dag}a_0 a_{\mu}\lambda\; \theta^{\mu},  \\
&& \textbf{K}_{B}= -\lambda^{\dag}a_0a_5 a_{\mu}\lambda\; \theta^{\mu}, \nonumber\\
&&\textbf{S}_{B} = -i\lambda^{\dag}a_0 a_{\mu}a_{\nu}\lambda \; \theta^{\mu}\wedge \theta^{\nu},\nonumber
\end{eqnarray} 
in which the set of $a_{\mu}$ matrices \cite{dharamboson}, read
\begin{eqnarray}
a_0 = i\left(\begin{array}{cc}
0 & \I \\ 
-\I & 0
\end{array}\right), \quad \boldsymbol{a} = i\left(\begin{array}{cc}
0 & \boldsymbol{\sigma} \\ 
\boldsymbol{\sigma} & 0
\end{array} \right), 
\end{eqnarray}
satisfying the following properties: $a^2_0=\I$, $\boldsymbol{a}^2=-\I$, and $\boldsymbol{a}^{\dag}=a_0\boldsymbol{a}a_0$. Moreover, they satisfy a very similar constitutive relation as the usual (fermionic) Clifford algebra. Nonetheless, the bosonic counterpart of the Clifford algebra is given by \cite{brauer}
\begin{eqnarray}\label{cliffboson}
\lbrace a_{\mu}, a_{\nu}\rbrace = 2g_{\mu\nu}\I.
\end{eqnarray}
and also they  do not commute with $\gamma_{\mu}$,
\begin{eqnarray}
[a_{\mu},\gamma_{\mu}] = 2g_{\mu\nu}i\gamma_5.
\end{eqnarray}
In addition, we can also define
\begin{eqnarray}
a_5 = \frac{i}{4!}\epsilon^{\mu\nu\alpha\beta}a_{\mu}a_{\nu}a_{\alpha}a_{\beta} = -\left(\begin{array}{cc}
\I & 0 \\ 
0 & -\I
\end{array} \right).
\end{eqnarray}
From these definitions, we have the following relations,
\begin{eqnarray}\label{a-gamma}
a_{\mu} = i\gamma_5 \gamma_{\mu}, \quad \gamma_{\mu} = i a_5 a_{\mu}.
\end{eqnarray} 
Notice the last above relations allow one to connect the fermionic with the bosonic representation of the Clifford algebra. The elements $\{ \theta^\mu \}$ are the dual basis of a given inertial-frame $\{ \textbf{e}_\mu \} = \left\{ \frac{\partial}{\partial x^\mu} \right\}$, with $\{x^\mu\}$ being the global space-time coordinates. 

So, the algebraic constraints presented in Eq.\eqref{bosonbilinear} reduce the possibilities of (only) six different classes (for which $\boldsymbol{J}$ is always non-null), giving rise to a  new spinor classification:
\begin{enumerate}
\item[$1_{B})$] $\sigma_{B}\neq0$, $\quad \omega_{B}\neq0$, \quad  $\textbf{K}_{B}\neq 0,$ \quad $\textbf{S}_{B}\neq0$;
\item[$2_{B})$] $\sigma_{B}\neq0$, $\quad \omega_{B}=0$, \quad $\textbf{K}_{B}\neq0,$ $\quad\textbf{S}_{B}\neq0$;
\item[$3_{B})$] $\sigma_{B}=0$, $\quad \omega_{B}\neq0$, \quad $\textbf{K}_{B}\neq0,$ $\quad\textbf{S}_{B}\neq0$;
\item[$4_{B})$] $\sigma_{B}=0=\omega_{B},$ \hspace{0.98cm}  $\textbf{K}_{B}\neq0,$ $\quad\textbf{S}_{B}\neq0$;
\item[$5_{B})$] $\sigma_{B}=0=\omega_{B},$ \hspace{0.98cm} $\textbf{K}_{B}=0,$ $\quad\textbf{S}_{B}\neq0$;
\item[$6_{B})$] $\sigma_{B}=0=\omega_{B},$ \hspace{0.98cm} $\textbf{K}_{B}\neq0,$ $\quad\textbf{S}_{B}=0$;
\end{enumerate}
with classes 1, 2 and 3 standing for regular spin-half fermions with bosonic traces and the remaining three holding singular classes. Thus, the classification above stands for the \emph{bosonic} counterpart of the Lounesto classification \cite{rodolfospinhalf} --- holding also some similar aspects. 

\section{Stage 1: On the \emph{regular} spinors framework}\label{stage1}
In this section we focus on investigating how to classify regular spinors by inspecting the phase factor $\alpha$.
To accomplish such a task, first consider a regular spinor, previously introduced in Eq.\eqref{set-1} and Eq.\eqref{set-2}. 
Accordingly, in the table below we scrutinize all the possibilities encoded on $\alpha_{R}$, in order to properly classify them, yielding the following relations\footnote{Notice the notation employed here: $\mathbb{C}$ stands for any complex number carrying a real and an imaginary part; $\mathbb{R}$ stands for any real number, and, finally, $Im$ represent a purely imaginary number.}
%\begin{strip}
\begin{table}[H]
\centering
\begin{tabular}{ccc}
\hline
\multicolumn{3}{c}{\textbf{Regular spinors}}\\
\hline 
\hline 
\;\;\;\;\;$\alpha_{R}$\;\;\;\;\; & \;\;\;\;\;Resulting Class\;\;\;\;\; & Constraints \\ 
\hline 
\hline 
  $\mathbb{C}$  & 1 & - \\
%\hline 
$\mathbb{R}$  &  2 & - \\ 
 %\hline 
 $Im$  & 3 & - \\ 
%\hline 
 $0$  & 6 & - \\ 
\hline 
\hline 
\end{tabular}
\caption{The phases constraints to classify  the new regular spinors.}
\end{table} 
%\end{strip}
\noindent In this way, we highlighted the requirements to determine a given class according to the respective values of $\alpha_{R}$. 
Imposing to the spinors in Eq.\eqref{set-1} and Eq.\eqref{set-2} to satisfy the \emph{bosonic} Dirac's equation counterpart, $(a_{\mu}p^{\mu}\mp m\I)\lambda_{j}(\textbf{p})=0$, automatically one find the relation: $\alpha_{R}=1$, thus, belonging to class 2 --- similarly to the fermionic case.

With the above results at hands, we should also analyse the possibility of summing regular spinors, looking towards obtain some informations about classes. That said, we can define the following relations
\begin{strip}
\begin{table}[H]
\centering
\begin{tabular}{ccc}
\hline
\:\: Resulting Class\:\: & \:\: $\lambda_{class\; k} = \lambda_{class\; i}+\lambda_{class\; j}$\:\: & \:\: Phases constraints\:\: \\ 
\hline 
\hline 
$1$ & $\lambda_{class 2}+ \lambda_{class 3}$ &  -
\vspace{0.1cm}\\\\ 
$1$ & $\lambda_{class 1}+ \lambda_{class 2}$ &  -
\vspace{0.1cm}\\\\
$1$ & $\lambda_{class 1}+ \lambda_{class 3}$ & -
\vspace{0.1cm}\\\\ 
$1$ & $\lambda_{class 1}+ \lambda_{class 1}$ & - 
\vspace{0.1cm}\\ 
$2$ & $\lambda_{class 2}+ \lambda_{class 2}$ & - 
\vspace{0.1cm}\\ 
$2$ & $\lambda_{class 2}+ \lambda_{class 2}$ &  $|a_{R_{i}}|^2 = |a_{R_{j}}|^2 =1$ (satisfying the \emph{bosonic} Dirac equation).
\vspace{0.1cm}\\\\ 
$3$ & $\lambda_{class 3}+ \lambda_{class 3}$ &  -
\vspace{0.1cm}\\
\hline 
\end{tabular} 
\label{tabelasomaregular}
\caption{Spinors combination for the regular sector.}
\end{table}
\end{strip}

\section{Stage 2: On the \emph{singular} spinors framework}\label{stage2}
In this section, we look towards define the structure of the singular spinors based on the new spin-half particles. Such a task is accomplished taking into account the Wigner's time-reversal operator, which in the spin-half representation, reads
\begin{equation}\label{thetawigner}
\Theta = \left(\begin{array}{cc}
0 & -1 \\ 
1 & \;\; 0
\end{array} \right),
\end{equation}
such an operator hold the following important relation $\Theta\vec{\sigma}\Theta^{-1} = -\vec{\sigma}^{*}$. Another interesting properties reads $\Theta^2 = -\I$ and $\Theta^{-1}=-\Theta$, given properties makes it possibles to define the singular counterpart of the spinors in \eqref{set-1} and \eqref{set-2}. A straightforward calculation, taking into account the aforementioned properties of the Wigner time-reversal operator, furnishes 
\begin{eqnarray}
\Theta\phi_{R/L}^{*}(\textbf{p})=\left(\I \mp \frac{\vec{\sigma}\cdot\vec{\textbf{p}}}{E+m}\right)\Theta\phi_{R/L}^{*}(\textbf{0}),
\end{eqnarray}
notice the action of the $\Theta$ operator in the last equation above, automatically flip the sign of the boost operator, when compared with the spinors \eqref{set-1} and \eqref{set-2} under action of the boost operator in \eqref{boostoperator}. 
Consequently, the summarized form of a singular structures can be displayed in the following fashion
\begin{eqnarray}
\lambda(\boldsymbol{p})=\sqrt{\frac{m}{2}}\left(\begin{array}{c}
\alpha_{S}\mathfrak{B}_{-}\Theta\phi_L^{*}(\textbf{0}) \\ 
\mathfrak{B}_{-}\phi_L(\textbf{0})
\end{array} \right), 
\end{eqnarray}
and/or
\begin{eqnarray}
 \lambda (\boldsymbol{p})=\sqrt{\frac{m}{2}}\left(\begin{array}{c}
\alpha_{S}\mathfrak{B}_{+}\phi_R(\textbf{0}) \\ 
\mathfrak{B}_{+}\Theta\phi_R^{*}(\textbf{0})
\end{array} \right).
\end{eqnarray}

In which we have defined the Lorentz boost operator as
\begin{equation}
\mathfrak{B}_{\pm}= \sqrt{\frac{E+m}{2m}}\left(\I \pm \frac{\vec{\sigma}\cdot\vec{\textbf{p}}}{E+m}\right).
\end{equation}
Where the rest frame singular spinors may be written as
{\footnotesize
\begin{eqnarray}
\lambda_1(\textbf{0}) = \sqrt{\frac{m}{2}}\left(\!\!\!\begin{array}{c}
\alpha_{S}\left(\!\!\!\begin{array}{c}
i \\ 
-1
\end{array}\!\!\!\right) \\\\ 
\quad\left(\!\!\!\begin{array}{c}
i \\ 
1
\end{array}\!\!\!\right)
\end{array}\!\!\!\right),\; \lambda_2(\textbf{0}) = \sqrt{\frac{m}{2}}\left(\!\!\!\begin{array}{c}
\alpha_{S}\left(\!\!\!\begin{array}{c}
i \\ 
1
\end{array}\!\!\!\right) \\ \\
\quad\left(\!\!\!\begin{array}{c}
-i \\ 
1
\end{array}\!\!\!\right)
\end{array}\!\!\!\right),\label{set-1.2}
\end{eqnarray}
}
and
{\footnotesize
\begin{eqnarray}
\lambda_3(\textbf{0})=\sqrt{\frac{m}{2}}\left(\!\!\!\begin{array}{c}
\alpha_{S}\left(\!\!\!\begin{array}{c}
-i \\ 
1
\end{array}\!\!\!\right) \\\\ 
\quad\left(\!\!\!\begin{array}{c}
i \\ 
1
\end{array}\!\!\!\right)
\end{array}\!\!\!\right),\;  \lambda_4(\textbf{0}) = \sqrt{\frac{m}{2}}\left(\!\!\!\begin{array}{c}
\alpha_{S}\left(\!\!\!\begin{array}{c}
-i \\ 
-1
\end{array}\!\!\!\right) \\\\ 
\quad\left(\!\!\!\begin{array}{c}
-i \\ 
1
\end{array}\!\!\!\right)
\end{array}\!\!\!\right). \label{set-2.2}  
\end{eqnarray}
}

Thus, a careful inspection of the phase factor yield 
%\begin{strip}
\begin{table}[H]
\centering
\begin{tabular}{ccc}
\hline
\multicolumn{3}{c}{\textbf{Singular spinors}}\\
\hline 
\;\;\;\;\;$\alpha_{S}$\;\;\;\;\; & \;\;\;\;Resulting Class\;\;\; & \;\;Constraints\;\; \\ 
\hline 
\hline 
 $\mathbb{C}$ & 4 & $|\alpha_{S}|^2\neq 1$ \\ 
%\hline 
 $\mathbb{C}$  & 5 & $|\alpha_{S}|^2= 1$ \\ 
%\hline 
$\mathbb{R}$  & 4 & $|\alpha_{S}|^2\neq 1$ \\
%\hline
 $\mathbb{R}$  & 5 & $|\alpha_{S}|^2=1$ \\
%\hline 
$Im$  & 4 & $|\alpha_{S}|^2\neq 1$ \\ 
%\hline 
$Im$  & 5 & $|\alpha_{S}|^2= 1$ \\ 
%\hline 
0  & 6 & - \\ 
%\hline 
\hline 
\hline 
\end{tabular} 
\caption{The phases constraints to classify singular spinors.}
\end{table}
%\end{strip}
\noindent With regard to the classes 4 and 5, we emphasize on the structure of the bilinear form $\boldsymbol{K}$. Such a bilinear form, in general grounds, depends on the modulo of the phase $a$. Hence, if $|\alpha_{S}|^2=1$, automatically $\boldsymbol{K}=0$, leading, then, the spinor to belong to class 5. Otherwise, if $|\alpha_{S}|^2\neq 1$, thus, it provides $\boldsymbol{K}\neq 0$, ensuring class 4. 

Another interesting fact is the right observation of to which class the sum of two distinct singular spinors belongs
\begin{strip}
\begin{table}[H]
\centering
\begin{tabular}{ccc}
\hline
\:\: Resulting Class\:\: & \:\: $\lambda_{class \; k} = \lambda_{class\; i}+\lambda_{class\; j}$\:\: & \:\: Phases constraints\:\: \\ 
\hline 
\hline 
$4$ & $\lambda_{class 4}+ \lambda_{class 4}$ & -
\vspace{0.1cm}\\
$4$ & $\lambda_{class 4}+ \lambda_{class 5}$ & -
\vspace{0.1cm}\\ 
$4$ & $\lambda_{class 4}+ \lambda_{class 6}$ & -
\vspace{0.1cm}\\\\ 
$4$ & $\lambda_{class 5}+ \lambda_{class 5}$ & -   
\vspace{0.1cm}\\
$4$ & $\lambda_{class 5}+ \lambda_{class 6}$ & -   
\vspace{0.1cm}\\\\ 
 $5$ & $\lambda_{class 5}+ \lambda_{class 5}$ & in which $|a_{S_{j}}|^2 =1$ and $|a_{S_{i}}|^2=1$ 
\vspace{0.1cm}\\\\ 
 $5$ & $\lambda_{class 4}+ \lambda_{class 4}$ & in which $|a_{S_{j}}|^2+|a_{S_{i}}|^2=2$ 
\vspace{0.1cm}\\\\
$6$ & $\lambda_{class 6}+ \lambda_{class 6}$ &  -
\vspace{0.1cm}\\\\ 
\hline 
\end{tabular} 
\label{tabelasomasingular}
\caption{Spinors combination for the singular sector}
\end{table} 
\end{strip}

Now we introduce the charge conjugation operator
\begin{eqnarray}
\mathcal{C} = \left(\begin{array}{cc}
0 & i\Theta \\ 
-i\Theta & 0
\end{array} \right)\mathcal{K},
\end{eqnarray}
where $\mathcal{K}$ complex conjugates to the right, Thus, under action of the charge-conjugation operator, the singular structure, in \eqref{set-1.2} and \eqref{set-2.2}, hold the following properties 
\begin{eqnarray}
&&\mathcal{C}\lambda_{j} = -\lambda_j, \quad\mbox{for j = 1, 2} \\
&&\mathcal{C}\lambda_{j} = +\lambda_j, \quad\mbox{for j = 3, 4}.
\end{eqnarray}
The above results are guaranteed under the following condition $\alpha_{S}=1$. Thus, accordingly the Tables above, such new structure belong to class-5.

\section{Final Remarks}
In the present essay, we introduced a complete set of regular and singular spin-half fermions holding some bosonic aspects, in its algebraic form, based on a previously defined structure. Moreover, taking into account such spinors, we explored all the phase constraints pointing towards the right definition of possible classes that such (unexpected) new particles may belong within an appropriated classification. Given a mathematical device lies on a careful analysis of the \emph{arbitrary} phase factor, namely $\alpha$, which may be represented by a real, complex, or a purely imaginary number. We must emphasize that since we are defining such spinors and their respective dual structures in terms of the bosonic representation of the Clifford algebra, we hope that theories constructed with such spinors will also inherit the same bosonic traces as the entities shown in Ref \cite{dharamboson}.

As can be seen, the regular spinors belonging to class-2 are a simulacrum of usual fermionic Dirac's spinors. Thus, for this specific case, by fixing $\beta=0$ in advance (according to the demand of locality) and also setting $ \alpha_{R} =1$, such particles with  bosonic traces obey a first-order equation and consequently provide a local theory \cite{dharamboson}. However, for the other classes, we cannot immediately state about the locality feature, without, at least, obtaining information about the quantum field operators and propagator structure. Moreover, as we know from the previous existing cases in the current literature, the locality feature can be later fixed by an appropriate dual structure redefinition. So far, what has been shown here is the possibility of populating all classes of the new spinor classification. Interestingly enough, we defined a new structure, belonging to the singular sector (more specifically, belonging to class-5), which are \emph{eigenspinors} of the charge-conjugation operator. Such structures can be understood as a hint towards spin-half fermions with bosonic traces holding similarity with Majorana and Elko fermions.

\section{Acknowledgements}
I am grateful to D. V. Ahluwalia for various discussions over the last months that culminated the completion of this work.

\bibliographystyle{unsrt}
\bibliography{refs}

\end{document}